\documentclass[fullpage]{article}
\usepackage{amsmath, amsthm, amssymb,bbm}
\DeclareMathAlphabet{\mathpzc}{OT1}{pzc}{m}{it}
\usepackage[pdftex]{hyperref}
\numberwithin{equation}{section}

\begin{document}

\begin{center}

{\bf\LARGE Light-Cone Superspace BPS Theory}\vskip 1 cm

{\bf \LARGE Patrick Hearin }\vskip 1cm
{\it Institute for Fundamental Theory,}
{\it Department of Physics, University of Florida,}
{\it Gainesville FL 32611, USA}\vskip 1cm
\end{center}
\vskip 2cm

\begin{abstract}
The BPS bound is formulated in light-cone superspace for the $\mathcal{N}=4$ superYang-Mills theory.  As a consequence of the superalgebra all momenta are shown to be expressed as a quadratic form in the relevant supertransformations, and these forms are used to derive the light-cone superspace BPS equations.  Finally, the superfield expressions are expanded out to component form, and the Wu-Yang Monopole boosted to the infinite momentum frame is shown to be a solution.\\\\\\
\textit{Keywords}: BPS; Light-Cone; Light-Front; Superspace; Superconformal Theories.
\end{abstract}

\newpage

\section{Introduction}

\ \ \ \ \ \ The $\mathcal{N}=4$ Yang-Mills theory has become one of the most interesting field theories, because it's perturbation theory is finite \cite{1}, \cite{2}.  The massless $\mathcal{N}=4$ theory is composed of a vector,  four Weyl fermions, and six real scalars.  All fields have global transformations under the conformal superalgebra $PSU(2,2|4)$, and local gauge transformations in the adjoint representation of the gauge group $G$.  

The method used in this article to obtain the massless $\mathcal{N}=4$ theory in four dimensions is dimensional reduction of a massless $\mathcal{N}=1$ vector multiplet in ten dimensions \cite{3}, \cite{4}, \cite{5}, \cite{6}.  The massless $\mathcal{N}=4$ multiplet's ten dimensional origin leaves a residue R-symmetry from the six reduced dimensions, and is the $SU(4)$ in $PSU(2,2|4)$.  

These reduced dimensions also have momentum in four dimensions, known as the central charges:  they commute with the four dimensional superPoincar\'e generators, and can be block diagonalized using Schur's lemma.  The diagonalized central charges can be used in massive theories to derive a lower bound on the mass.  This phenomenon is known as the Bogomol'nyi, Prasad, Sommerfield bound, or BPS bound \cite{7}, \cite{8}, \cite{9}, \cite{10}, \cite{11}.  This bound can be derived algebraically by choosing the fields to be in their rest frame in the supersymmetric algebra.  When some amount of the block diagonalized central charge's two eigenvalues are equal to the mass the bound is saturated;  accordingly, when saturation occurs some of the conjugate supercharge operators annihilate the Clifford vacuum.  When all of the charges are equal to the mass this is the half multiplet, because it has the same number of states as the massless theory;  consequently, the field theory formulation of the half multiplet can be derived using the Higgs Mechanism, because the total number of states does not change when the Higgs Mechanism is applied. 

To derive the BPS bound in the field theory the Higgs Mechanism breaks the massless symmetries, the fermions are set to zero, and the potential is minimized.  The assumptions made pertaining to fermions and potential are obviously not required to derive the bound since the algebra holds non-perturbetivly for all fields and potentials, but they make the factorization of the Hamiltonian easier.  Using this simplified theory the algebraic logic for the mass bound is reproduced using the bosonic Hamiltonian.  Saturation of the bound is given by the field configurations satisfying the first order Bogomol'nyi differential equations \cite{7}, \cite{8}.  When saturation occurs the field configurations describe the half multiplet by preserving half of the supersymmetry.  In this article, the light-cone equivalent of these bosonic Bogomol'nyi equations will be generalized to superspace;  furthermore, this field theory derivation is analogous to the algebra since it makes no superfluous assumptions.  To derive these supersymmetric BPS equations the light-cone superfield formalism is used as the main tool.

The light-cone formalism is an extremely powerful method for understanding on-shell supersymmetric field theories, and was originally proposed by Dirac \cite{12}.  A concise review of the conventions can be found in appendix \ref{sec:a}.  The light-cone technique exploits the equations of motion to eliminate the non-physical fields from the theory;  thus, for a gauge theory in four dimensions the vector particle is reduced by two fields.  The non-physical fermionic degrees of freedom are eliminated by projecting the equations of motion.  This light-cone projection yields a unique anti-symmetry in the supersymmetric algebra when gauge interactions are included.  Half of the transformations, known as the kinematical transformations, remain linear in the interacting theory, and the other projection, known as the dynamical transformations, has non-linear interactions.  In superspace this anti-symmetry of the algebra makes the light-cone field theory an optimal place to derive new theories from algebraic first principles, because a theory is totally determined by the dynamical transformations: \cite{13}, \cite{14}, \cite{15}.

The light-cone superalgebra leads to an algebraic formulation of the light-cone Hamiltonian for maximally supersymmetric theories as a product of conjugate dynamical transformations in superspace.  This quadratic form was first discovered in \cite{13} for the $N=4$ theory.  Now the quadratic form has been formulated for the three dimensional BLG theory \cite{14}, \cite{15}, \cite{16} and $\mathcal{N}=8$ supergravity theory to order $\kappa^2$ in gravitational coupling \cite{17}.  In this article, the superalgebra is used to show that all momenta and central charges can be expressed as quadratic forms for the $\mathcal{N}=4$ superYang-Mills theory. 

These quadratic forms are useful to formulate the BPS theory in light-cone superspace, since they are related to anti-commutators that are positive definite.  After the massless symmetries are broken by a boundary condition in terms of the light-cone spatial variables the same expression used to derive the algebraic saturation is used to formulate the BPS bound in the field theory by replacing the operators with infinitesimal transformations.  This is the first superfield generalization of the Bogomol'nyi equations and these equations describe a superfield moduli space that is the supermanifold generalization of Atiyah and Hitchin Manifold \cite{18}. Finally, the superfield expressions are expanded to component form, and the bosonic solutions of these equations are shown to be the infinite momentum frame boost of the Wu-Yang monopole.  These solutions imply that light-cone solitons are the infinite momentum frame boost of the equal-time solutions.

\section{Dimensional Reduction}
\label{sec:2}

\ \ \ \ In this section the dimensional reduction techniques will be briefly reviewed, and the four dimensional conserved supercharges will be derived.  The ten dimensional $\mathcal{N}=1$ theory is composed of a massless vector particle $A^{a}_M$ and a thirty-two component spinor $\lambda^a$.  These fields transform under supersymmetry, the Poincar\'e group $\mathbb{R}^{9,1} \ltimes SO(9,1)$, and the adjoint representation of the gauge group $G$, with structure constant $f^{abc}$.  The lower case letters $a,b,c,d,e$ denote the adjoint representation's indices, and $M, N$ are the ten dimensional space-time indices.  

In ten dimensions the vector particle has eight degrees of freedom; consequently, twenty-four fermionic degrees of freedom must be eliminated out of the original thirty-two components to achieve supersymmetry.  The method used to achieve statistical equality is imposing the Majorana and Weyl constraints on the spinor  
\begin{align}&\bar{\lambda}^a=\lambda^{aT} C,\label{eq:constraint 1}\\ \nonumber \\
&\Gamma_{11}\lambda^a=\lambda^a,\label{eq:constraint 2}\end{align} 
where $\bar{\lambda}^a=\lambda^{a\dag}\Gamma^0$.  These constraints eliminate sixteen fermionic degrees of freedom, and with the equations of motion they make the number of fermions equal to the number of bosons.\footnote{The representation used to define the constraints \eqref{eq:constraint 1}, \eqref{eq:constraint 2} is reviewed in appendix \ref{sec:a}.}

Furthermore, the Poincar\'e, gauge invariant Lagrangian
\begin{equation}\mathcal{L}=-\frac{1}{4}F^a_{MN}F^{aMN}+i\frac{1}{2}\bar{\lambda}^a\Gamma_M D^{abM}\lambda^b,\end{equation}
where 
\begin{align}&D^{ab}_M=\partial_M\delta^{ab}-gf^{abc}A^{c}_M,\\ \nonumber \\
&F^{a}_{MN}=\partial_MA^a_N-\partial_NA^a_M+gf^{abc}A^b_MA^c_N,\end{align}
was shown in \cite{3} to be supersymmetric when the Majorana and Weyl constraints are imposed under the transformations
\begin{align}&\delta A^a_M=i\bar{\alpha} \Gamma_M \lambda^a+D^{ab}_M\omega^b,\label{eq:transformation 1}\\ \nonumber \\
&\delta \lambda^a=\frac{1}{4}\Gamma^{MN}F^a_{MN}\alpha+f^{abc}\lambda^b\omega^c,\label{eq:transformation 2}\end{align}
where $\alpha$ is a 32 component constant spinor that parameterizes the global supersymmetric transformations, $\omega^a$ is the local gauge parameter, and $\Gamma^{MN}=[\Gamma^M,\Gamma^N]$. The previous supersymmetric transformations \eqref{eq:transformation 1}, \eqref{eq:transformation 2} have the gauge symmetry added to them to make the gauge and supersymmetry compatible, and is discussed in \cite{6}.  The conserved current for the supersymmetric transformations is the sum of the original supersymmetric current plus the gauge current.  Both currents are conserved when the equations of motion 
\begin{align}&D^{ab}_MF^{bMN}=-\frac{i}{2}f^{abc}\bar{\lambda}^b\Gamma^N\lambda^c,\\ \nonumber \\
&\Gamma_MD^{abM}\lambda^b=0,\tag{2.20b}\end{align}
are used, so the constants in front of the currents are arbitrary;  therefore, the current
\begin{equation}\bar{\alpha} J_M=-\frac{1}{4}F^a_{PQ}\bar{\alpha}\Gamma^{PQ}\Gamma_M\lambda^a-iF^{a}_{MN}D^{abN}\omega^b+i\frac{1}{2}gf^{abc}\bar{\lambda}^a\Gamma_M\lambda^b\omega^c,\label{eq:current}\end{equation}
is conserved when the equations of motion are used.

Now the $\mathcal{N}=4$ theory's algebra in four dimensions will be reduced from its $\mathcal{N}=1$ ten dimensional origin.  The ten dimensional Poincar\'e algebra dimensionally reduces to $SO(9,1)\supset SO(3,1)\otimes SO(6) $, where the four-dimensional space-time group $SO(3,1)$ is accompanied by a $SO(6)\sim SU(4)$ R-symmetry, and four-dimensional translations.  Before reducing the ten dimensional $\mathcal{N}=1$ superalgebra 
\begin{equation}\{Q_\alpha,\bar{Q}_\beta\}=(\Gamma_M)_{\alpha\beta}P^M,\label{eq:super algebra}\end{equation} 
it is projected with the matrices
\begin{align}&\Pi^\pm=-\frac{1}{2}\Gamma^\mp\Gamma^\pm,\\ \nonumber \\
&\Pi^\pm Q_{\alpha}=Q_{\alpha\pm},\ \ \ \bar{Q}_\alpha\Pi^\pm=\bar{Q}_{\alpha\mp}.\end{align}
Thus, applying the Majorana, Weyl constraints, and projecting the spinors yields
\begin{align}
\bar{Q}_+ &=-i\big (0,0,0,\bar{q}_1,...,0,0,0,\bar{q}_4,q^1,0,0,0,...,q^4,0,0,0\big ),\label{eq:plus} \\ \nonumber \\
\bar{Q}_-&=i\big(0,0,\bar{\mathcal{Q}}_1,...,0,0,\bar{\mathcal{Q}}_4,0,0,\mathcal{Q}^1,0,0,0,...,\mathcal{Q}^4,0,0\big ),\label{eq:minus} \end{align}
and leaves four Grassmann variables for each projection that transform globally under the fundamental representation of $SU(4)$ $(\mathbf{4},\mathbf{\bar{4}})$, with spinor indices $m,n=1,2,3,4$.  The plus projection \eqref{eq:plus} contains the kinematical charges $q^m$ and it's complex conjugate.  While, the minus projection \eqref{eq:minus} contains the dynamical charges $\mathcal{Q}^m$ and it's conjugate, where in four dimensions operators that are dynamical are denoted by capital calligraphic letters.  When these supercharges are substituted into \eqref{eq:super algebra} this yields the algebra
\begin{align}&\{q^m,\bar{q}_n\}=\sqrt{2}\delta^m_nP^+,\label{eq:momenta 1}\\ \nonumber \\
&\{\bar{\mathcal{Q}}_m,\mathcal{Q}^n\}=\sqrt{2}\delta^m_n\mathcal{P}^-,\label{eq:momenta 2}\\ \nonumber \\
&\{q^m,\bar{\mathcal{Q}}_n\}=\sqrt{2}\delta^m_n P,\label{eq:momenta 3}\\ \nonumber \\
&\{q^m,\mathcal{Q}^n\}=\sqrt{2}\mathcal{Z}^{mn},\label{eq:central charge}\end{align}
where $P^+=\frac{1}{\sqrt{2}}(P^0+P^3)$, $\mathcal{P}^-=\frac{1}{\sqrt{2}}(P^0-P^3)$, $P=\frac{1}{\sqrt{2}}(P^1+iP^2)$.  

Dimensional reduction makes the six extra dimensional momentum constant real numbers, or the central charge.  The central charges \eqref{eq:central charge} commute with the  superPoincar\'e algebra after reduction to four dimensions.  These charges transform under the anti-symmetric representation of $SU(4)$, $\bold{6}$, and are defined by the matrix $\mathcal{Z}^{mn}=\frac{1}{\sqrt{2}}\Sigma^{mnI}P^I$, where $I=4,5,6,7,8,9$ and the matrix $\Sigma^{mnI}$ is defined in appendix \ref{sec:a}:  this operator has six real components, since it obeys the duality condition
\begin{equation}\mathcal{Z}^{mn}=\frac{1}{2}\epsilon^{mnpq}\bar{\mathcal{Z}}_{pq}.\end{equation}

Now the field theory is reduced to four dimensions.  First, the unphysical degrees of freedom: $A^{a+},\ A^{a-},\ \lambda^a_-$ are eliminated;  thus, the temporal light-cone coordinate field $A^{a+}$ is chosen to be equal to zero, or the light-cone gauge condition, and the equations of motion are used to solve for $A^{a-}$,  $\lambda^a_-$.\footnote{The light-cone method for eliminating the unphysical degrees of freedom can be reviewed in \cite{6}.}  When the light-cone gauge condition is chosen $A^{a+}=0$ this constrains the gauge parameter to be independent of the $x^-$ coordinate $\delta A^{a+}=\partial^+\omega^a=0$, and due to the supersymmetric variation \eqref{eq:transformation 1} the gauge parameter can be solved for \footnote{The gauge parameter \eqref{eq:parameter} was originally derived in \cite{6}.}
\begin{equation}\omega^a=-i\bar{\alpha}_+\Gamma^+\frac{1}{\partial^+}\lambda^a_+.\label{eq:parameter}\end{equation}
Now all of the unphysical degrees of freedom have been eliminated leaving the massless little group in ten dimensions symmetry $SO(8)\subset SO(9,1)$;  consequently, the dimensional reduction $SO(8)\supset SO(2)\otimes SO(6)$ reduces the bosonic field $A^a_K$, where $K=1,2,4,5,6,7,8,9$, transforming under $SO(8)$ into: a four-dimensional vector field transforming under $U(1)$ helicity, $A^a=A^a_1+iA^a_2$, and scalars defined by the matrix $C^{amn}=\frac{1}{\sqrt{2}}\Sigma^{mnI}A^{aI}$.  The scalars transform under the $\bold{6}$ of $SU(4)$, and have the duality $C^{amn}=\frac{1}{2}\epsilon^{mnpq}\bar{C}_{pq}^a$ that leaves six real components. The left over spinor
\begin{align}\bar{\lambda}_+=-i\big (\bar{\chi}^a_1,0,0,0
,...,
\bar{\chi}^a_4,0,0,0,
0,0,0,\chi^{a1},...,
0,0,0,\chi^{a4}\big ),\label{eq:spinor}\end{align}
is Majorana, Weyl and the $\chi^{am}$ transform under: $U(1)$ helicity, the fundamental representation of $SU(4)$, $\bold{4}$, and the $\bar{\chi}^a_m$ transform under $\bar{\bold{4}}$.   The results for the Lagrangian, and the supertransformation reduction can be found in: \cite{5}, \cite{6} respectively.  

The only quantity needed for this article is the four dimensional light-cone supercharges, and they are not found in the previous literature. To reduce the supercharge the supersymmetry parameter's projections $\alpha=\alpha_++\alpha_-$ \footnote{The supersymmetric parameters projections $\alpha_+$, $\alpha_-$ yield two different parameters upon reduction, called $\alpha^m$, $\beta^m$ in \cite{6}.  For the purposes of this article, these parameters are set equal to each other $\alpha^m=\beta^m$, without loss of generality;  furthermore, the supersymmetric parameter $\alpha$ is rescaled by a factor of $-\sqrt{2}$ when reduced, or the four four dimensional parameters inside of $\alpha$ are $-\frac{1}{\sqrt{2}}\alpha^m$ and $-\frac{1}{\sqrt{2}}\bar{\alpha}_m$.  This is the same rescaling that is used in \cite{6} to match superspace expressions.} divide the plus component of the current \eqref{eq:current} into $\alpha_-$ multiplying the kinematical charge, and $\alpha_+$ multiplying the dynamical.\footnote{The technique that divides the kinematical from the dynamical symmetries was originally used in \cite{6}.}  Thus, the kinematical charges are \footnote{Field theory expressions are denoted by $\mathbbm{q}$, $\mathbbm{Q}$, $\mathbbm{P}$.}
\begin{equation}\mathbbm{q}^m=2i\int d^3x\{\partial^+C^{amn}\bar{\chi}^a_n-\partial^+A^a\chi^{am}\}.\label{eq:kinematical charge}\end{equation}
Using the matrix identities 
\begin{align}&\Gamma\bar{\Gamma}\Gamma^+=-i\gamma\bar{\gamma}\gamma^+\otimes \bold{1}_8=2i\left(\begin{array}{cc}0 & \sigma^+ \\ 0 & 0\end{array}\right)\otimes \bold{1}_8, \\ \nonumber \\
&\bar{\Gamma}\Gamma\Gamma^+=-i\bar{\gamma}\gamma\gamma^+\otimes \bold{1}_8=2i\left(\begin{array}{cc}0 & 0 \\ \sigma^- & 0\end{array}\right)\otimes \bold{1}_8, \\ \nonumber \\
&\Gamma\Gamma^I\Gamma^+=-\sqrt{2}i\left(\begin{array}{cc}\sigma &0  \\0 & 0\end{array}\right)\otimes\left(\begin{array}{cc}0 & \Sigma^{Imn} \\ \Sigma_{Imn}& 0\end{array}\right), \\ \nonumber \\ 
&\bar{\Gamma}\Gamma^I\Gamma^+=-\sqrt{2}i\left(\begin{array}{cc}0 &0  \\0 & \bar{\sigma}\end{array}\right)\otimes \left(\begin{array}{cc}0 & \Sigma^{Imn} \\ \Sigma_{Imn}& 0\end{array}\right), \\ \nonumber \\
&\Gamma^I\Gamma^J\Gamma^+=-i\left(\begin{array}{cc}0 &\sigma^-  \\ \sigma^+ & 0\end{array}\right)\otimes \left(\begin{array}{cc}\Sigma^{Imp}\Sigma^{J}_{pn} & 0\\ 0& \Sigma^{I}_{mp}\Sigma^{Jpn}\end{array}\right),\end{align}
where $\sigma^\pm=\frac{1}{\sqrt{2}}(\bold{1}_2\pm\sigma^3)$, $\sigma=\frac{1}{\sqrt{2}}(\sigma^1+i\sigma^2)$, and the identity $\bar{C}^a_{mn}\partial^+C^{bnp}\frac{1}{\partial^+}\bar{\chi}^a_p=-\frac{1}{2}(\bar{C}^a_{mn}C^{bnp}\bar{\chi}^c_p+\frac{1}{2}\bar{C}^a_{np}\partial^+C^{bnp}\frac{1}{\partial^+}\bar{\chi}^a_m)$ the dynamical charges reduce to
\begin{align}\mathbb{Q}^m=&2i\int d^3x\{\partial C^{amn}\bar{\chi}^a_{n}-\partial\bar{A}^a\bar{\chi}^{am}-gf^{abc}(\partial^+ \bar{A}^{a}A^{b}\frac{1}{\partial^+}\chi^{cm}-\nonumber \\
& -\bar{A}^aC^{bmn}\bar{\chi}^c_n+C^{amn}\partial^+\bar{C}^{b}_{np}\frac{1}{\partial^+}\chi^{cp}+i\frac{1}{\sqrt{2}}\bar{\chi}^a_m\chi^b\frac{1}{\partial^+}\chi^{cs})\}.\label{eq:dynamical charge}\end{align}
The supercharge reduction to four dimensions is the only new result in this section.  Now that all four-dimensional quantities have been derived the superspace expressions can be formulated.

\section{SuperSpace}
\label{sec:3}

\ \ \ \  To formulate the superspace theory the Grassmann variables $\theta^m$ are defined to transform under the fundamental representation of $SU(4)$, $\bold{4}$, and the complex conjugate $\bar{\theta}_m$ transforms under $\bar{\bold{4}}$.  These coordinates have an anti-commuting derivative $\partial^m=\frac{\partial}{\partial\bar{\theta}_m}$ that has the complex conjugation $(\partial^m)^*=-\bar{\partial}_m$, and satisfies
\begin{equation}\{\partial^m,\bar{\theta}_n\}=\delta^m_n;\end{equation}
accordingly, the usual space-time can be redefined such that the whole theory can be placed in superspace.  The new coordinate is defined by
\begin{equation}y=(x,\bar{x},x^+, y^-= x^--\frac{i}{\sqrt{2}}\theta^m\bar{\theta}_m),\end{equation}
and all fields are placed inside a chiral superfield
\begin{align}\phi^a(y)=&\frac{1}{\partial^+}A^a(y)+\frac{i}{\sqrt{2}}\theta^m\theta^n\bar{C}^a_{mn}(y)+\frac{i}{\partial^+}\theta^m\bar{\chi}^a_m(y)+\nonumber\\ \nonumber \\
&+\frac{\sqrt{2}}{6}\epsilon_{mnpq}\theta^m\theta^n\theta^p\chi^q(y)+\frac{1}{12}\epsilon_{mnpq}\theta^m\theta^n\theta^p\theta^q\partial^+\bar{A}^a(y).\end{align}
The transformations on the superfield can be expressed in terms of the charge operators in superspace for the free theory
\begin{align}&q^m=-\partial^m+i\frac{1}{\sqrt{2}}\theta^m\partial^+,\label{eq:kinematical transformation}\\ \nonumber \\
&\mathcal{Q}^m=\frac{\bar{\partial}}{\partial^+}q^m.\label{eq:dynamical charge operator}\end{align}
Taking a different linear combination of the anti-commuting derivative and Grassmann variable yields the chiral derivative
\begin{equation}d^m=-\partial^m-i\frac{1}{\sqrt{2}}\theta^m\partial^+.\end{equation}
For a massless free theory the supercharge and chiral derivative satisfy 
\begin{align}&\{q^m,\bar{q}_n\}=\sqrt{2}i\delta^m_n\partial^+,& &\{\mathcal{Q}^m,\bar{\mathcal{Q}}_n\}=\sqrt{2}i\delta^m_n\frac{\partial\bar{\partial}}{\partial^+},& &\{d^m,\bar{d}_n\}=-\sqrt{2}i\delta^m_n\partial^+,& &\{d^m,q^n\}=0,&\label{eq:anti-commutation}\nonumber \\ \nonumber \\ \end{align}
and the chiral derivative annihilates the chiral superfield
\begin{equation}d^m\phi^a=0;\end{equation}
thus, a supersymmetric transformation preserves chirality for the light-cone formalism.  Finally, there is the inside-out constraint which follows from the duality of the scalars
\begin{equation}d^md^n\bar{\phi}^a=\frac{1}{2}\epsilon^{mnpq}\bar{d}_p\bar{d}_q\phi^a,\end{equation}
and it allows the superfield to be conjugated with covariant derivatives
\begin{equation}\bar{\phi}^a=\frac{1}{48}\bar{d}^4\frac{1}{\partial^{+2}}\phi^a,\end{equation}
where $d^4=\epsilon_{mnpq}d^md^nd^pd^q$ and $\bar{d}^4=\epsilon^{mnpq}\bar{d}_m\bar{d}_n\bar{d}_p\bar{d}_q$.  The technique for reducing to the component theory can be reviewed in \cite{6}.\footnote{For a complete review of light-cone superspace the reader is also referred to \cite{5}, \cite{13}.}

Originally, in \cite{13} the light-cone Hamiltonian was shown to be a quadratic form
\begin{equation}\mathbb{P}^-=i\frac{1}{\sqrt{2}}\int d^3xd^4\theta d^4\bar{\theta}\{\Delta^m\bar{\phi}^a\frac{1}{\partial^+}\bar{\Delta}_{m}\phi^a\},\end{equation}
and the non-linear infinitesimal dynamical transformation 
\begin{equation}\bar{\Delta}_{m}\phi^a=\frac{1}{\partial^+}(\partial\delta^{ab}-f^{abc}\partial^+\phi^c)\bar{q}_m\phi^b,\label{eq:dynamical transformation}\end{equation}
was derived using algebraic first principles.\footnote{The dynamical transformation \eqref{eq:dynamical transformation} is a covariant derivative for the residual gauge symmetry transformation, this phenomenon is discussed in \cite{6}.}  Using the equations \eqref{eq:kinematical charge}, \eqref{eq:dynamical charge} superspace expressions \footnote{The equations \eqref{eq:super kinematical charge}, \eqref{eq:super dynamical charge} superspace forms where derived for the $\mathcal{N}=2$ theory coupled to a Wess-Zumino multiplet \cite{19}.}
\begin{align}
&\bar{\mathbbm{q}}_{m}=2\int d^4xd^4\theta d^4\bar{\theta}\{\bar{\phi}^a\frac{1}{\partial^{+}}\bar{q}_{m}\phi^a\},\label{eq:super kinematical charge}\\ \nonumber\\
&\bar{\mathbb{Q}}_{m}=2\int d^4xd^4\theta d^4\bar{\theta}\{\bar{\phi}^a\frac{\partial}{\partial^{+2}}\bar{q}_{m}\phi^a-\frac{2}{3}gf^{abc}\bar{\phi}^a\frac{1}{\partial^{+2}}(\bar{q}_m\phi^b\partial^+\phi^c)\},\label{eq:super dynamical charge}
\end{align}
 and the superalgebra it can be shown that all momenta are quadratic forms, because in Hilbert spaces the infinitesimal change of the conserved charge $\mathbb{F}$ due to another transformation $\mathbb{G}$ is their algebraic commutator $[\mathcal{F},\mathcal{G}]$ 
\begin{equation}i\delta_{\mathbb{G}}\mathbb{F}=[\mathcal{G},\mathcal{F}].\end{equation}
Using the previous equation all momenta come from the infinitesimal variation of the associated charges due to the supersymmetric algebra \eqref{eq:momenta 1}, \eqref{eq:momenta 2}, \eqref{eq:momenta 3}, \eqref{eq:central charge} respectively:
\begin{align}
&\mathbb{P}^+\alpha^m\bar{\alpha}_m=\frac{1}{\sqrt{2}}\alpha^m\bar{\alpha}_n\{\bar{q}_m,q^n\} =\frac{1}{\sqrt{2}}\alpha^m[\bar{\alpha}_nq^n,\bar{q}_m]=\frac{1}{\sqrt{2}}i\alpha^m\delta_{\bar{\alpha}_nq^n} \bar{\mathbbm{q}}_m,\label{eq:ac1}\\
&\mathbb{P}^-\alpha^m\bar{\alpha}_m =\frac{1}{\sqrt{2}}\alpha^m\bar{\alpha}_n\{\bar{\mathcal{Q}}_m,\mathcal{Q}^n\}=\frac{1}{\sqrt{2}}\alpha^m[\bar{\alpha}_n \mathcal{Q}^n,\bar{\mathcal{Q}}_m]=\frac{1}{\sqrt{2}}i\alpha^m\delta_{\bar{\alpha}_n\Delta^n} \bar{\mathbb{Q}}_m,\label{eq:ac2}\\
&\mathbb{P}\bar{\alpha}_m\alpha^m=\frac{1}{\sqrt{2}}\bar{\alpha}_m\alpha^n\{\bar{q^m,\mathcal{Q}}_n,\} =\frac{1}{\sqrt{2}}\bar{\alpha}_m[\alpha^n\bar{\mathcal{Q}}_n,q^m]=\frac{1}{\sqrt{2}}i\bar{\alpha}_m\delta_{\alpha^n\bar{\Delta}_n}\mathbbm{q}^m,\label{eq:ac3}\\
&\mathbb{Z}^{mn}\bar{\alpha}_m\bar{\alpha}_n=\frac{1}{\sqrt{2}}\bar{\alpha}_m\bar{\alpha}_n\{q^m,\mathcal{Q}^n\} =\frac{1}{\sqrt{2}}\bar{\alpha}_m[\bar{\alpha}_n\mathcal{Q}^n,q^m]=\frac{1}{\sqrt{2}}i\bar{\alpha}_m\delta_{\bar{\alpha}_n\Delta^n}\mathbbm{q}^m,
\label{eq:ac4}\end{align}
where the notation $\delta_{\bar{\alpha}_n\Delta^n}$, $\delta_{\bar{\alpha}_nq^n}$ denotes the infinitesimal variation with respect to the operator in the subscript.  For example, the central charge quadratic form can be derived from the algebra by varying the kinematical charge \eqref{eq:super kinematical charge} with the dynamical transformation \eqref{eq:dynamical transformation}
\begin{align}&\mathbb{Z}^{mn}\bar{\alpha}_m\bar{\alpha}_n=\frac{1}{\sqrt{2}}i\bar{\alpha}_m\delta_{\bar{\alpha}_n\Delta^n}\mathbbm{q}^m= \nonumber \\ \nonumber \\&=\sqrt{2}i\bar{\alpha}_m\bar{\alpha}_n\int d^4xd^4\theta d^4\bar{\theta}\{\Delta^n\bar{\phi}^a\frac{1}{\partial^+}q^m\phi^a-\bar{\phi}^a\frac{1}{\partial^+}\Delta^nq^m\phi^a\}.\end{align}
Integrating by parts and using the inside-out constraint yields the quadratic form
\begin{equation}\mathbb{Z}^{mn}=2\sqrt{2}i\int d^3xd^4\theta d^4\bar{\theta}\{q^m\bar{\phi}^a\frac{1}{\partial^+}\Delta^n\phi^a\}.\end{equation}
Varying the dynamical charge \eqref{eq:super dynamical charge} with the kinematical transformation \eqref{eq:kinematical transformation} also yields the previous result. The other momenta's quadratic forms also follow from the light-cone supersymmetric algebra, and their derivations are similar to the central charge:

\begin{align} 
&\delta^m_n\mathbb{P}^-=2\sqrt{2}i\int d^3xd^4\theta d^4\bar{\theta}\{\Delta^m\bar{\phi}^a\frac{1}{\partial^{+}}\bar{\Delta}_n \phi^a\},\label{eq:quadratic form 1}\\ \
&\delta^m_n\mathbb{P}^+=2\sqrt{2}i\int d^3xd^4\theta d^4\bar{\theta}\{\delta^m\bar{\phi}^a\frac{1}{\partial^+}\bar{\delta}_{n}\phi^a\},\label{eq:quadratic form 2}\\
&\delta^m_n\mathbb{P}=2\sqrt{2}i\int d^3xd^4\theta d^4\bar{\theta}\{\delta^m\bar{\phi}^a\frac{1}{\partial^+}\bar{\Delta}_n\phi^a\},\label{eq:quadratic form 3}\\
&\mathbb{Z}^{mn}=2\sqrt{2}i\int d^3xd^4\theta d^4\bar{\theta}\{\delta^m\bar{\phi}^a\frac{1}{\partial^+}\Delta^n\phi^a\},\label{eq:quadratic form 4}
\end{align}
where the infinitesimal kinematical transformations (without the supersymmetric parameter) are denoted by $\delta^m\phi^a=q^m\phi^a$, $\bar{\delta}_m\phi^a=\bar{q}_m\phi^a$.  Calculating these quadratic forms by varying the conserved charges proves that the longitudinal quadratic forms are positive definite, because they are equal to the anti-commutators that are complex conjugates of each other.  The off diagonal terms in \eqref{eq:quadratic form 1}, \eqref{eq:quadratic form 2}, \eqref{eq:quadratic form 3} are trivially zero in this construction, because the anti-commutators \eqref{eq:momenta 1}, \eqref{eq:momenta 2}, \eqref{eq:momenta 3} are zero for off diagonal charges.  The central charge quadratic form is also antisymmetric $\mathbb{Z}^{mn}=-\mathbb{Z}^{nm}$, because the kinematical charges $q^m$, $q^n$ in the quadratic form can be integrated by parts to switch the indices $m,n$.  This antisymmetry can also be comprehended by observing that the component form of \eqref{eq:quadratic form 4} indices are just the scalar field's indices $C^{amn}$ which is discussed in section \ref{sec:4}.  Now that the quadratic forms have been formulated, the BPS theory will be derived using them.  The method for deriving the equations \eqref{eq:quadratic form 1}, \eqref{eq:quadratic form 2}, \eqref{eq:quadratic form 3}, \eqref{eq:quadratic form 4} is new, and much simpler then algebraic first principle techniques.  The quadratic form expressions for the central charge, plus component, and transverse momenta are also all new results.
\section{Light-Cone BPS}

\subsection{Representations}

\ \ \ \ In this section the half-multiplet will be formulated for the light-cone superalgebra.  Representations can be found by placing particles in their relevant reference frames, and using the superalgebra to infer which supercharges have non-zero anti-commutation relations.  These non-zero anti-commutations define the different multiplets by acting with conjugated supercharge operators on the Clifford vacuum $|\Omega>$.  The Clifford vacuum is defined by the fact that non-conjugate supercharges annihilate it.  Since, the Clifford vacuum is an irreducible representation of the superPoncar\'e algebra it has helicity or spin depending on the particles mass.  For an algebra with $\mathcal{N}$ supersymmetry operators in four dimensions a massless Clifford vacuum has helicity $U(1)$, and eigenvalue $h$.  Choosing the massless particles to be in either reference frame $P_\mu=(E,0,0,\pm E)$ forces half the conjugate supercharges to annihilate the Clifford vacuum.  The remaining conjugate supercharges generate the massless multiplet by acting on the Clifford vacuum state with the greatest or least helicity, yielding $2^\mathcal{N} $ states. 

 Massive multiplets have more states then the massless, because placing the particles in the rest frame does not force any of the conjugate supercharges to annihilate the Clifford vacuum that has $(2s+1)$ states under the $SU(2)$ spin group with third component of spin eigenvalue $s$;  furthermore, since none of the conjugate supercharges annihilate the Clifford vacuum the block diagonalized central charge eigenvalues do not vanish.  Thus, for $s=0$, the largest massive representation has no central charges creating saturation, and is called the long multiplet, it has $2^{2\mathcal{N}}$ states.  The smaller multiplets are called short multiplets, and are defined by how many of the $\frac{\mathcal{N}}{2}$ central charge eigenvalues saturate the BPS bound.  When $n$ of the central charge eigenvalues saturate the bound, this generates  $2^{2(\mathcal{N}-n)}$ states.  Thus, for the $\mathcal{N}=4$ theory the number of states for the different multiplets is: massless 16, half multiplet 16, quarter multiplet 64, and the long multiplet 256.

For a massless representation in the light-cone formalism, the particle can be chosen to travel along the positive or negative z-axis
\begin{align}&(P^+=\sqrt{2}E,\mathcal{P}^-=0,P=0,\bar{P}=0),\\ \nonumber \\
&(P^+=0,\mathcal{P}^-=\sqrt{2}E,P=0,\bar{P}=0).\end{align}
Both reference frames take the central charge \eqref{eq:central charge} to zero, because either the kinematical or dynamical conjugated charge operators annihilate the Clifford vacuum.  Thus, the massless multiplet is generated by acting on the Clifford vacuum states with greatest or least helicity with either the kinematical or dynamical conjugate charges depending on the choice of the frame
\begin{align}\zeta^{+m}=\frac{\mathcal{Q}^m}{2^\frac{1}{4}\sqrt{E}}\ \ \ \ or\ \ \ \ \zeta^{-m}=\frac{q^m}{2^\frac{1}{4}\sqrt{E}}.\label{eq:conjugate operators}\end{align}
These operators satisfy two different Clifford algebras
\begin{align}\{\zeta^{\pm m},\bar{\zeta}^{\pm}_n\}=\delta^m_n,\ \ \ \ \{\zeta^{\pm m},\bar{\zeta}^{\mp}_n\}=\{\zeta^{\pm m},\zeta^{\pm n}\}=\{\zeta^{\pm m},\zeta^{\mp n}\}=0\end{align}
and the plus or minus denotes raising and lower of the helicity by the transverse rotation operator \cite{13}
\begin{align}&j=x\bar{\partial}-\bar{x}\partial+\frac{1}{2}(\theta^m\bar{\partial}_m-\bar{\theta}_m\partial^m)-\frac{i}{4\sqrt{2}}[d^m,\bar{d}_m]\frac{1}{\partial^+},\\ \nonumber \\ 
&[j,\bar{q}_m]=-\frac{1}{2}\bar{q}_m,\ \ \ \ \ \ [j,\bar{\mathcal{Q}}_m]=\frac{1}{2}\bar{\mathcal{Q}}_m.\end{align}
Acting on the states with helicity $h=\pm1$ with the conjugate of \eqref{eq:conjugate operators} generates 16 massless states:
\begin{align}&|\Omega_{h=\pm1}>& &one\ \ helicity=\pm1,&\nonumber \\ \nonumber \\  
&\bar{\zeta}^\mp_m|\Omega_{h=\pm1}>& &four\ \ helicity=\pm\frac{1}{2},&\nonumber\\ \nonumber \\
&\bar{\zeta}^\mp_m\bar{\zeta}^\mp_n|\Omega_{h=\pm1}>& &six\ \ helicity=0,&\nonumber\\ \nonumber \\
&\bar{\zeta}^\mp_m\bar{\zeta}^\mp_n\bar{\zeta}^\mp_p|\Omega_{h=\pm1}>& &four\ \ helicity=\mp\frac{1}{2},&\nonumber\\ \nonumber \\
&\bar{\zeta}^\mp_1\bar{\zeta}^\mp_2\bar{\zeta}^\mp_3\bar{\zeta}^\mp_4|\Omega_{h=\pm1}>& &one\ \ helicity=\mp1.&\end{align}

To define massive theories we must consider the central charge, because the rest frame does not force any supercharges to annihilate the Clifford vacuum.  After applying Schur's lemma the block diagonal central charge is composed of two complex eigenvalues $\mathcal{Z}_n$, where: $\mathcal{Z}_1=\mathcal{Z}_2$, $\mathcal{Z}_3=\mathcal{Z}_4$; thus, the algebra takes the form:
\begin{align}&\{q^{m},\bar{q}_{n}\}=\sqrt{2}\delta^m_nP^+,\\ \nonumber \\
&\{\mathcal{Q}^{m},\bar{\mathcal{Q}}_{n}\}=\sqrt{2}\delta^m_n\mathcal{P}^-,\\ \nonumber \\
&\{q^{m},\mathcal{Q}^{n}\}=\sqrt{2}\epsilon^{mn}\mathcal{Z}_n,\label{eq:diagonalized central charge}\end{align}
where the $n$ in \eqref{eq:diagonalized central charge} is not summed over, and the matrix $\epsilon^{mn}$ is a two index anti-symmetric matrix 
\begin{equation}\epsilon^{mn}=\left(\begin{array}{cccc}0 & 1 & 0 & 0 \\ -1 & 0 & 0 & 0 \\0 & 0 & 0 & 1 \\ 0 & 0 & -1 & 0\end{array}\right).\end{equation}
Dividing the identity $\mathbf{1}_4=\mathbf{1}^++\mathbf{1}^-$ and the anti-symmetric matrix $\epsilon=\epsilon^++\epsilon^-$ \footnote{The matrices $\sigma^\pm=\frac{1}{\sqrt{2}}(\sigma_0\pm \sigma_3)$ and $\sigma=\frac{1}{\sqrt{2}}(\sigma_1+i\sigma_2)$ are discussed in section \ref{sec:a}.}
\begin{align}&\mathbf{1}^+=\frac{1}{\sqrt{2}}\sigma^+\otimes \mathbf{1}_2,\ \ \ &\mathbf{1}^-=\frac{1}{\sqrt{2}}\sigma^-\otimes \mathbf{1}_2,\\ \nonumber \\
&\epsilon^+=i\frac{1}{\sqrt{2}}\sigma\otimes\sigma^2,\ \ \ &\epsilon^-=i\frac{1}{\sqrt{2}}\sigma^\dag\otimes\sigma^2,\end{align}
allows the kinematical and dynamical charges to be arranged into two operators
\begin{align}&\mathpzc{a}^m=(\mathbf{1}^+)^m_n\mathcal{Q}^{n}-(\epsilon^+)^{mn}\bar{q}_{n}+(\mathbf{1}^-)^m_n q^n-(\epsilon^-)^{mn}\bar{\mathcal{Q}}_n,\label{eq:a}\\ \nonumber \\
&\mathpzc{b}^m=(\mathbf{1}^+)^m_n\mathcal{Q}^{n}+(\epsilon^+)^{mn}\bar{q}_{n}+(\mathbf{1}^-)^m_n q^n+(\epsilon^-)^{mn}\bar{\mathcal{Q}}_n.\label{eq:b}\end{align}
In the light-cone coordinate rest-frame
\begin{equation}(P^+=\frac{1}{\sqrt{2}}M,\mathcal{P}^-=\frac{1}{\sqrt{2}}M,P=0,\bar{P}=0),\end{equation}
these operators satisfy the algebra
\begin{align}&\{\mathpzc{a}^m,\mathpzc{a}^n\}=\{\mathpzc{b}^m,\mathpzc{b}^n\}=\{\mathpzc{a}^m,\bar{\mathpzc{b}}_{n}\}=0,\\ \nonumber \\
&\{\mathpzc{a}^m,\bar{\mathpzc{a}}_{n}\}=2\delta^m_n(M+\sqrt{2}Re(\mathcal{Z}_n)), \\ \nonumber \\
&\{\mathpzc{b}^m,\mathpzc{\bar{b}}_{n}\}=2\delta^m_n(M-\sqrt{2}Re(\mathcal{Z}_n)). \label{eq:b anti-commutator}\end{align}
The anti-commutations $\{\mathpzc{a}^m,\mathpzc{\bar{a}}_{n}\}$, $\{\mathpzc{b}^m,\mathpzc{\bar{b}}_{n}\}$ are positive definite for $m=n$;  thus, the mass is bounded below by the central charge eigenvalues for each $m$	
\begin{equation}M\geq\sqrt{2}Re(\mathcal{Z}_m).\label{eq:BPS bound}\end{equation}
The different multiplets are generated when the operators $\bar{b}_m$ annihilate the Clifford vacuum with spin $s$;  since, there are only two central charges, annihilation of the vacuum by $\bar{\mathpzc{b}}_1$ or $\bar{\mathpzc{b}}_2$ sets the first eigenvalue to the mass, and annihilation by $\bar{\mathpzc{b}}_3$ or $\bar{\mathpzc{b}}_4$  sets the second eigenvalue to the mass.  The half multiplet is defined by all the $\bar{\mathpzc{b}}_m$ annihilating the Clifford vacuum state with $s=0$
\begin{equation}\bar{\mathpzc{b}}_m|\Omega_{s=0}>=0,  \end{equation}
and all the central charges are equal to the mass
\begin{equation}M=\sqrt{2}Re(\mathcal{Z}_m).\end{equation}   
To generate the half multiplet the operator \eqref{eq:a} is divided into two sets depending on whether they raise or lower the third component of spin
\begin{align}
\bar{\mathpzc{a}}^+_1&=\bar{\mathpzc{a}}_1,\ \ \ \bar{\mathpzc{a}}^+_2 =\bar{\mathpzc{a}}_2,\\ \nonumber \\
\bar{\mathpzc{a}}^-_1&=\bar{\mathpzc{a}}_3,\ \ \ \bar{\mathpzc{a}}^-_2=\bar{\mathpzc{a}}_4,
\end{align}
where the plus and minus denote this raising and lowering of the third component of spin: $[j,\bar{\mathpzc{a}}^+_i]=\frac{1}{2}\bar{\mathpzc{a}}^+_i$, and $[j,\bar{\mathpzc{a}}^-_i]=-\frac{1}{2}\bar{\mathpzc{a}}^-_i$, where $i=1,2$.  Thus, the operators 
\begin{equation}\eta^{\pm }_i=\frac{1}{2\sqrt{M}}\mathpzc{a}^{\pm }_i,\end{equation}
satisfy two Clifford algebras 
\begin{align}\{\eta^{\pm }_i,\bar{\eta}^{\pm }_j\}=\delta_{ij},\ \ \ \ \{\eta^{\pm }_i,\bar{\eta}^{\mp }_j\}=0,\ \ \ \ \{\eta^{\pm }_i,\eta^{\pm }_j\}=0,\ \ \ \ \{\eta^{\pm }_i,\eta^{\mp }_j\}=0,\end{align} 
and acting on the Clifford vacuum with $\eta^{\pm }_i$ yields:
\begin{align}&|\Omega_{s=0}>& &one\ \ s=0,&\nonumber\\ \nonumber \\
&\bar{\eta}^\pm_i|\Omega_{s=0}> & &four\ \ s=\pm\frac{1}{2},&\nonumber\\ \nonumber \\
&\bar{\eta}^\pm_1\bar{\eta}^\pm_2|\Omega_{s=0}>& &two\ \ s=\pm 1,&\nonumber\\ \nonumber \\
&\bar{\eta}^\pm_i\bar{\eta}^\mp_j|\Omega_{s=0}>& &four\ \ s=0,&\nonumber\\ \nonumber \\
&\bar{\eta}^\pm_i\bar{\eta}^\pm_j\bar{\eta}^\mp_k|\Omega_{s=0}>& &four\ \ s=\pm\frac{1}{2},&\nonumber\\ \nonumber \\
&\bar{\eta}^+_1\bar{\eta}^-_2\bar{\eta}^+_2\bar{\eta}^-_1|\Omega_{s=0}>& &one\ \ s=0.&\end{align}
This multiplet has the same field content as a massless multiplet; thus, in the field theory this mutiplet can be obtained by using the Higgs mechanism on the massless $\mathcal{N}=4$ field theory.  The algebraic formulation of the half multiplet will be used to derive the BPS bound \eqref{eq:BPS bound} in the field theory based purely on algebraic first principles and no superfluous assumptions, because the bound can be derived based on preserving half the supersymmetries using \eqref{eq:a}, \eqref{eq:b}.

\subsection{Mass Factorization}

\label{sec:3.2}

\ \ \ \ In this section the BPS bound will be derived for the $\mathcal{N}=4$ field theory in superspace.  The BPS bound found from the algebra \eqref{eq:BPS bound} holds for all the particles in the theory;  consequently, the field theory must have a factorization of the mass that holds for all particles.  To achieve this factorization the massless symmetries are broken by constant scalar vacuum configurations that minimize the potential 
\begin{align}\frac{1}{16}g^2f^{abc}f^{ade}[C^{bmn}C^{cpq}\bar{C}^d_{mn}\bar{C}^e_{pq}+\frac{1}{\partial^+}(C^{bmn}\partial^+\bar{C}^{c}_{mn})\frac{1}{\partial^+}(C^{dpq}\partial^+\bar{C}^e_{pq})]. \end{align}
This potential has non-local terms that do not change the minimum from zero, because these terms are squared.  For any constant scalar field these non-local terms vanish;  therefore, the moduli space for this theory is described by the conditions 
\begin{align}f^{abc}C^{bmn}C^{cpq}=0&, &f^{abc}C^{bmn}\partial^+\bar{C}^c_{mn}=0,\label{eq:potential min}\end{align}
which is the normal $\mathcal{N}=4$ orbifold 
\begin{equation}\mathbb{R}^{6r}/ \mathcal{W},\end{equation}
where $r$ is the rank of the gauge group $G$ and $\mathcal{W}$ is its Weyl group \cite{20}.  At a non-singular point in the previous orbifold the gauge group is broken to the Cartan torus $U(1)^r$ and the R-symmetry is broken to $SO(5)\sim Sp(4)$ by the constant Higgs field 
\begin{equation}C=\nu\cdot H,\label{eq:Higgs}\end{equation}
where $H$ is the Cartan sub-algebra, $\nu$ is a vector of real numbers with length $r$, and the root basis $\beta_l$ is chosen such that $\nu\cdot\beta_l\neq 0$.  This scheme yields $r$ massless multiplets and as many massive multiplets as the number of root generators \cite{21}.
 
To calculate the BPS bound the spontaneous symmetry breaking is achieved by a spatial boundary condition imposed on the scalar fields.  This boundary condition requires the energy be finite on the boundary, or the theory must map to the vacuum; accordingly, the field theory on the boundary must reduce to the gauge vacuum manifold parameterized by the Higgs field \eqref{eq:Higgs}.  For an equal-time theory this boundary is $S^2$, and the invariant deformations of the gauge vacuum manifold to the boundary yield topological conservation laws characterized by the second homotopy group of the gauge vacuum manifold $\pi_2(G/U(1)^r)=\pi_1(U(1)^r)=\mathbb{Z}^r$.  

For our theory these finite energy solutions, and topological conservation laws are in the light-cone coordinate system, or on the spatial boundary $x^+=-x^-$, which is an equal-time foliation of space.  Since, the light-cone coordinate system is on the light-front, or constant $x^+$, this equal-time foliation needs to be boosted to the infinite momentum frame.  To boost to the infinite momentum frame the static equal-time boundary is boosted along the third axis, and then the rapidity is approximated around $\frac{\pi}{4}$.  This process yields a finite boundary in terms of $x^1,x^2,x^-$ that spontaneously breaks light-cone theories in the same manner as equal-time theories discussed in the previous paragraph.  For a broken light-cone theory the topological solitons are the infinite momentum frame boost of the equal-time solutions;  thus, when this boundary condition is imposed it yields a massive theory that has a BPS bound. 

To derive the BPS bound the block diagonal central charge anti-commutations \eqref{eq:diagonalized central charge} are used instead of the other basis \eqref{eq:central charge} to diagonalize the quadratic form
\begin{equation}\epsilon^{mn}\mathbb{Z}_n=2\sqrt{2}i\int d^4xd^4\theta d^4\bar{\theta}\{\delta^{m}\bar{\phi}^a\frac{1}{\partial^+}\Delta^{n}\phi^a\}.\label{eq:diagonalized quadratic form}\end{equation}
Just like the algebra \eqref{eq:a}, \eqref{eq:b} the kinematical and dynamical variations are divided
\begin{align}&\Diamond^m=(\mathbf{1}^+)^m_n\Delta^{n}-(\epsilon^+)^{mn}\bar{\delta}_{n}+(\mathbf{1}^-)^m_n \delta^n-(\epsilon^-)^{mn}\bar{\Delta}_n,\\ \nonumber \\
& \nabla^m=(\mathbf{1}^+)^m_n\Delta^{n}+(\epsilon^+)^{mn}\bar{\delta}_{n}+(\mathbf{1}^-)^m_n \delta^n+(\epsilon^-)^{mn}\bar{\Delta}_n,\label{eq:nabla}\end{align}
where these are the infinitesimal version of the operators $\mathpzc{a},\mathpzc{b}$ respectively. These infinitesimal transformations have an associated conserved charge
\begin{equation}\mathbbm{b}^m=(\mathbf{1}^+)^m_n \mathbb{Q}^n+(\epsilon^+)^{mn}\bar{\mathbbm{q}}_n+(\mathbf{1}^-)^m_n\mathbbm{q}^{n} +(\epsilon^-)^{mn}\bar{\mathbb{Q}}_n .\end{equation}
Varying this conserved charge with \eqref{eq:nabla} 
\begin{equation}\alpha^n[\bar{\alpha}_m \mathpzc{b}^m,\bar{\mathpzc{b}}_n]=i\alpha^n(\delta_{\bar{\alpha}_m\nabla^m} \bar{\mathbbm{b}}_n),\end{equation}
yields the field theory equivalent of the operator expression \eqref{eq:b anti-commutator}, a positive semi-definite quantity.  After integrating by parts and using the inside out constraint the final expression in the rest frame is
\begin{align}\{\mathpzc{b}^m,\bar{\mathpzc{b}}_{n}\}=2\sqrt{2}i\int d^43 d^4\theta d^4\bar{\theta}\{\nabla^m\bar{\phi}^a\frac{1}{\partial^+}\bar{\nabla}_{n}\phi^a\}=2\delta^m_n(M-\sqrt{2}Re(\mathbb{Z}_{n})). \end{align}
The surface terms that come from integration by parts in the previous equation are zero, because they depend inversely on the radius.  The non-zero terms are charges that only depend on angular variables on the boundary, and all these terms are contained in the central charge quadratic form;  therefore, the superfield equivalent of the Bogomol'nyi technique is
\begin{equation}M\delta^m_n=\sqrt{2}\int d^3x d^4\theta d^4\bar{\theta}\{\nabla^m\bar{\phi}^a\frac{1}{\partial^+}\bar{\nabla}_{n}\phi^a\}+ \delta^m_n\sqrt{2}Re(\mathbb{Z}_n),\end{equation}
and when m=n the quadratic form is positive semi-definite making the BPS bound
\begin{equation}M\geq \sqrt{2}Re(\mathbb{Z}_m),\label{eq:superspace bound}\end{equation}
valid field theoretically for all fields. Saturation is achieved when the equation
\begin{equation}\bar{\nabla}_{m}\phi^a=0,\label{eq:Superfield BPS}\end{equation}
is satisfied.  The other infinitesimal transformations yield the mass
\begin{equation}M= \sqrt{2}Re(\mathbb{Z}_m)=\frac{1}{\sqrt{2}}\int d^3x d^4\theta d^4\bar{\theta}\{\Diamond^{m}\bar{\phi}^a\frac{1}{\partial^+}\bar{\Diamond}_{m}\phi^a\}.\end{equation}
These superspace equations have a simple interpretation in terms of the component theory which is derived in the next section.

\subsection{Component Equations}

\label{sec:4}

\ \ \ \ Now the component theory is derived from superspace to compare with conventional calculations with the fermions and potential set to zero for a $SU(2)$ gauge theory.  This component theory's equations are equivalent to dimensionally reducing the ten dimensional theory with the fermions and potential set to zero.  The dimensional reduction leaves a four dimensional Georgi-Glashow model with globally $SU(4)$ symmetric scalars $C^{amn}$.

In the field theory the choice of the minus component of the momentum as the energy forces the vacuum to be described by the kinematical charge taking the superfield to zero.  This fact can be seen through the quadratic forms \eqref{eq:quadratic form 1}, \eqref{eq:quadratic form 2}, \eqref{eq:quadratic form 3}, \eqref{eq:quadratic form 4} vanishing when
\begin{equation}\bar{\delta}_m \phi^a=\bar{q}_m\phi^a=0.\label{eq:component vacuum}\end{equation}
Expanding out the previous equation to component form yields
\begin{equation}\bar{C}^{a}_{mn}=\chi^a =\partial^+A^a=0.\end{equation}
Therefore, states with no momentum are given by vector fields that are independent of $x^-$, and all other fields are zero.  Although, the vacuum has to be Lorentz invariant making $A^a=0$, and \eqref{eq:component vacuum} describes the trivial vacuum configuration, or all fields are zero.

For massless particles the only choice for the reference frame is along the positive third axis, because the choice along the negative axis yields the vacuum \eqref{eq:component vacuum}.  Thus, this choice forces the dynamical transformations to take the superfield to zero.  This makes the longitudinal $\mathbb{P}^-$ momentum \eqref{eq:quadratic form 2}, and the transverse $\mathbb{P},\ \bar{\mathbb{P}}$ momentum \eqref{eq:quadratic form 3} vanish when the dynamical symmetry annihilates the superfield
\begin{equation}\bar{\Delta}_m\phi^a=\frac{1}{\partial^+}[(\partial\delta^{ab}-f^{abc}\partial^+\phi^c)\bar{q}_m\phi^b]=0,\end{equation}
leaving $\mathbb{P}^+$ arbitrary.   Expanding out the previous equations with the fermions equal to zero yields two equations
\begin{align}&\mathcal{D}^{ab}\bar{C}^b_{rm}=0,\label{eq:massless equation 1}\\ \nonumber \\
&\frac{1}{\partial^+}(\mathcal{D}^{ab}\partial^+A^b)=\frac{1}{4}gf^{abc}\frac{1}{\partial^+}(\partial^+\bar{C}^b_{mn}C^{c\ mn}),\label{eq:massless equation 2}\end{align}
where $\mathcal{D}^{ab}=\partial \delta^{ab}-gf^{abc}A^c$ is the transverse covariant derivative left over from the residual gauge invariance.  The equations \eqref{eq:massless equation 1}, \eqref{eq:massless equation 2} can be derived from dimensionally reducing the ten dimensional light-cone Hamiltonian from the stress energy tenser
\begin{equation}T^{MN}=-F^{aMP}F^{aN}_P+g^{MN}\mathcal{L}.\end{equation}
The light-cone Hamiltonian is the plus, minus component of the stress energy tenser
\begin{equation}T^{+-}=\frac{1}{2}(F^{a+-}F^{a+-}-F^aF^a)+\frac{1}{2}F_{iI}^{a}F_{iI}^{a}+\frac{1}{4}F_{IJ}^{a}F_{IJ}^{a},\end{equation}
and the equations \eqref{eq:massless equation 1}, \eqref{eq:massless equation 2} are found by setting a subset of the components of the field strength tenser to zero
\begin{equation}F^{a+-}\pm F^a=F^a_{iI}=F^a_{IJ}=0.\end{equation}
Dimensional reduction of the previous equations yields the scalar potential is zero
\begin{equation}\Sigma^{Imn}\Sigma^J_{pq}F^a_{IJ}=2f^{abc}C^{bmn}\bar{C}^c_{pq}=0,\end{equation}
and the scalar is a covariant constant
\begin{equation}\Sigma^I_{mn}(F^a_{1I}+iF^a_{2I})=\sqrt{2}\mathcal{D}^{ab}\bar{C}^b_{mn}=0,\end{equation}
or equation \eqref{eq:massless equation 1}.  The vector field equation \eqref{eq:massless equation 2} is found by choosing the light-cone gauge $A^{a+}=0$, and substituting in the dependent field
\begin{equation}A^{a-}=\frac{1}{\partial^{+2}}[\mathcal{D}^{ab}\partial^+\bar{A}^b+\mathcal{\bar{D}}^{ab}\partial^+A^b-\frac{1}{2}gf^{abc}(\partial^+\bar{C}^b_{mn}C^{c\ mn})],\label{eq:equation of motion}\end{equation}
into the equation
\begin{equation}F^{a+-} + F^a=2\frac{1}{\partial^+}(\mathcal{D}^{ab}\partial^+\bar{A}^b)-\frac{1}{2}gf^{abc}\frac{1}{\partial^+}(\partial^+\bar{C}^b_{mn}C^{c\ mn})=0,\end{equation}
where $F^a=\frac{1}{\partial^+}(\mathcal{D}^{ab}\partial^+\bar{A}^b-\bar{\mathcal{D}}^{ab}\partial^+A^b)$.

The solutions to \eqref{eq:massless equation 1}, \eqref{eq:massless equation 2} are a transverse pure gauge vector field
\begin{equation}A=U^\dag\partial U,\end{equation} 
and scalars that are covariant constants in the transverse space
\begin{equation}C^{amn}=K^{mn}e^{\int dx \{U^\dag\bar{\partial} U\}},\end{equation}
where $U$ is an unitary $3\times3$ matrix that is an arbitrary function of space-time, and $K^{mn}$ is a $3\times3$ matrix that is an arbitrary function of the longitudinal coordinate.

Massive theories have field configurations satisfying the superfield equation \eqref{eq:Superfield BPS}. Using the inside-out constraint \eqref{eq:Superfield BPS} becomes
\begin{align}&(\delta^m+\epsilon^{mn}\bar{\Delta}_n)\phi^a=\nonumber \\ \nonumber \\
&=q^m\phi^a+\epsilon^{mn}\frac{1}{\partial^+}(\partial\delta^{ab}-f^{abc}\partial^+\phi^c)\bar{q}_m\phi^b=0.\label{eq:superfield BPS expanded}\end{align}
The solutions to these equations have mass equal to the central charge eigenvalues
\begin{align}M=\frac{1}{\sqrt{2}}Re(\mathbb{Z}_1+\mathbb{Z}_3).\end{align}
The eigenvalues can be derived from the component form of \eqref{eq:diagonalized quadratic form} with the identity $C^{amp}\partial^+C^{bnq}\bar{C}^c_{pq}=\frac{1}{4}C^{amn}\partial^+C^{bpq}\bar{C}^c_{pq}$, and the fermions equal to zero

\begin{align}\mathbb{Z}^{mn}=&2[\partial^+A^a\bar{\mathcal{D}}^{ab}C^{bmn}-\partial^+C^{amn}\frac{1}{\partial^+}(\bar{\mathcal{D}}^{ab}\partial^+A^b)+\nonumber\\ \nonumber \\& +\frac{1}{4}gf^{abc}\partial^+C^{amn}\frac{1}{\partial^+}(\partial^+\bar{C}^b_{pq}C^{cpq}))].\label{eq:central charge component form}\end{align}
Contracting the $SU(4)$ indices with $\epsilon_{mn}$ yields the central charge eigenvalues

\begin{align}
&M=\frac{1}{\sqrt{2}}Re(\mathbb{Z}_1+\mathbb{Z}_3)=\frac{1}{2\sqrt{2}}\epsilon_{mn}Re(\mathbb{Z}^{mn})=\nonumber\\ \nonumber \\
&=\frac{1}{2\sqrt{2}}[\partial^+A^a\bar{\mathcal{D}}^{ab}\Phi^{b}+\partial^+\bar{A}^a\mathcal{D}^{ab}\Phi^{b}-\partial^+\Phi^{a}\partial^+A^{a-}],\end{align}
where $\Phi^a=\epsilon^{mn}\bar{C}^a_{mn}$.  Integrating by parts yields the $U(1)$ light-cone electric charge plus the equation of motion \eqref{eq:equation of motion} with the potential minimized by the conditions \eqref{eq:potential min}

\begin{align}&M=\frac{1}{\sqrt{2}}\nu \int d^3x\{(\bar{\partial}(\partial^+A^3)+\partial(\partial^+\bar{A}^3)-\partial^+(\partial^+A^{3-})\}=\nonumber\\ \nonumber \\
&=\frac{1}{\sqrt{2}}\nu\int d^3x\{\partial_\nu F^{3+\nu}\}=\frac{1}{\sqrt{2}}\nu Q^{LC}_E,
\label{eq:charge}\end{align}
where on the boundary $\Phi^a=2\nu\delta^{a3}$ as discussed in \ref{sec:3.2}. 

Expanding out the equation \eqref{eq:superfield BPS expanded} with the fermions equal to zero yields
\begin{align} &\partial^+A^a=\frac{1}{4}\epsilon^{mn}\mathcal{D}^{ab}\bar{C}^b_{mn},\label{eq: BPS component 1}\\ \nonumber\\
&-\frac{1}{4}\epsilon_{mn}\partial^+C^{amn}=\frac{1}{\partial^+}(\mathcal{D}^{ab}\partial^+\bar{A}^b)+\frac{1}{4}gf^{abc}\frac{1}{\partial^+}(\bar{C}^b_{mn}\partial^+C^{c mn}).\label{eq: BPS component 2}\end{align}
These equations have the same interpretation as \eqref{eq:massless equation 1}, \eqref{eq:massless equation 2} from the stress energy tenser.  The mass in the rest frame and fermions/potential set to zero can be factorized with an arbitrary four by four anti-symmetric complex matrix $X^{mn}$
\begin{align}M&=\frac{1}{\sqrt{2}}(T^{+-}+T^{++})=\nonumber\\ \nonumber \\
&=\frac{1}{\sqrt{2}}[\frac{1}{2}(F^{a+-}F^{a+-}-F^{a}F^{a})+F^{a+i}F^{a+i}+\frac{1}{4}D^{abi}\bar{C}^b_{mn}D^{aci}C^{cmn}+\frac{1}{2}D^{ab+}\bar{C}^b_{mn}D^{ac+}C^{cmn}+\nonumber\\ \nonumber \\
&+\frac{1}{16}g^2f^{abc}f^{ade}[C^{bmn}C^{cpq}\bar{C}^d_{mn}\bar{C}^e_{pq}]=\nonumber\\ \nonumber \\
&=\frac{1}{\sqrt{2}}[\frac{1}{2}|(X^{mn}( F^{a+-}+F^a)+D^{+ ab}C^{bmn})|^2+|(X^{mn} F^{a+i}-\frac{1}{2}D^{iab}C^{bmn})|^2]+\nonumber\\ \nonumber \\
&+\frac{1}{16}g^2f^{abc}f^{ade}C^{bmn}C^{cpq}\bar{C}^d_{mn}\bar{C}^e_{pq}-\frac{1}{2}(\bar{X}_{mn}F^{a+-}D^{+ ab}C^{bmn}+X^{mn}F^{a+-}D^{+ ab}\bar{C}^b_{mn})+\nonumber\\ \nonumber \\
&+\frac{1}{2}(\bar{X}_{mn} F^{a+i}D^{iab}C^{bmn}+X^{mn}F^{a+i}D^{iab}\bar{C}_{mn}^b)],\label{eq:LC Mass}\end{align}
where the matrix $X^{mn}$ satisfies $\bar{X}_{mn}X^{mn}=1$, and $X^{mn}\bar{C}^a_{mn}$ has to be real.  These conditions do not specify the anti-symmetric matrix $\epsilon^{mn}$ for $X^{mn}$, but $X^{mn}=\frac{1}{2}\epsilon^{mn}$ satisfies the required identities.  Thus, setting $X^{mn}=\frac{1}{2}\epsilon^{mn}$ the last terms in \eqref{eq:LC Mass} become the electric charge by using the equation of motion \eqref{eq:equation of motion}, and the bound

\begin{equation}M\geq \frac{1}{\sqrt{2}}\nu Q^{LC}_E,\end{equation}  
agrees with \eqref{eq:charge}, and is saturated by the equations:

\begin{align} &F^{a+i}=\frac{1}{4}D^{iab}\Phi^b \label{eq:light-cone coordinate BPS 1},\\ \nonumber \\
&F^{a+-}+F^a=-\frac{1}{2}D^{+ ab}\Phi^{b},\label{eq:light-cone coordinate BPS 2}\\ \nonumber \\
&f^{abc}C^{bmn}C^{cpq}=0,\label{eq:light-cone coordinate BPS 3}\\ \nonumber \\
&f^{abc}C^{bmn}\partial^+\bar{C}^c_{mn}=0.\label{eq:light-cone coordinate BPS 4}\end{align}
The equations \eqref{eq:light-cone coordinate BPS 1}, \eqref{eq:light-cone coordinate BPS 2} agree with \eqref{eq: BPS component 1}, \eqref{eq: BPS component 2} after choosing the light-cone gauge and substituting the dependent vector field;  furthermore, the equations \eqref{eq:light-cone coordinate BPS 3}, \eqref{eq:light-cone coordinate BPS 4} are the potential minimization conditions \eqref{eq:potential min}.

The equations \eqref{eq:light-cone coordinate BPS 1}, \eqref{eq:light-cone coordinate BPS 2}, \eqref{eq:light-cone coordinate BPS 3} have the same solutions in terms of the 't Hooft Polyakov monopole \cite{22}, \cite{23}, because they are just linear combinations of the equal time BPS equations.  To comprehend how these equations have 't Hooft Polyakov monopole solution the ten dimensional mass in the rest frame with fermions/potential equal to zero is reduced, and factorized in the equal-time coordinate system yielding

\begin{align}M=&\int d^3x\{\frac{1}{2}(B^{ax}B^{ax}+E^{ax}E^{ax})+\frac{1}{4}(D^{abx}\bar{C}^b_{mn}D^{acx}C^{cmn}+D^{ab0}\bar{C}^b_{mn}D^{ac0}C^{cmn})+\nonumber \\ \nonumber \\ \nonumber
&+\frac{1}{4}g^2f^{abc}f^{ade}C^{bmn}C^{cpq}\bar{C}^d_{mn}\bar{C}^e_{pq}\}=\\ \nonumber \\ \nonumber
&=\int d^3x\{|X^{mn}B^{ax}-\frac{1}{\sqrt{2}}ae^{i\theta}D^{abx}C^{amn}|^2+|X^{mn}E^{ax}-\frac{1}{\sqrt{2}}iae^{i\kappa}D^{abx}C^{amn}|^2\}+\\ \nonumber \\ 
&+a\sin(\kappa)Q_E+b\cos(\theta)Q_B,\end{align} 
where $\kappa$, $\theta$ are phases, and $a$, $b$ are arbitrary real numbers such that $a^2+b^2=1$.  Thus, saturation of the bound
\begin{equation}M \geq a\sin(\kappa)Q_E+b\cos(\theta)Q_B,\end{equation}
occurs when

\begin{align}&B^{ax}=\frac{1}{2\sqrt{2}}ae^{i\theta}D^{abx}\Phi^{a},\label{eq:ET BPS equation 1}\\ \nonumber \\
&E^{ax}=\frac{1}{2\sqrt{2}}iae^{i\kappa}D^{abx}\Phi^a,\label{eq:ET BPS equation 2}\\ \nonumber \\
&D^{ab0}\Phi^b=0,\\ \nonumber \\
&f^{abc}C^{bmn}C^{cpq}=0, \end{align}
and the solutions are foliated in the equal-time coordinate or static

\begin{align}&\Phi^a=\frac{r^a}{r^2}(\zeta coth(\zeta)-1),\label{eq:ET solution 1}\\ \nonumber \\
&A^{a0}=a ie^{i\theta}\frac{r^a}{r^2}(\zeta coth(\zeta)-1),\label{eq:ET solution 2}\\ \nonumber \\
&A^{ax}=b e^{i\kappa}\epsilon^{axy}\frac{r^y}{r^2}(1-\zeta csch(\zeta)),\label{eq:ET solution 3}\end{align}
where $\zeta=\nu g r$, and $r$ is the three dimensional radius.\footnote{Mass factorizations in equal-time theories can be reviewed in \cite{10}.}  Therefore, since the field strength satisfies
\begin{align}&\frac{1}{2}i(ae^{i\theta}-be^{i\kappa})\mathcal{D}^{ab}\Phi^b=\frac{1}{\sqrt{2}}(E^{a}-iB^{a})=\frac{1}{\sqrt{2}}(F^{a+1}+iF^{a+2}),\\ \nonumber \\
&-i(ae^{i\theta}-be^{i\kappa})D^{3 ab}\Phi^b=-E^{a3}+iB^{a3}=F^{a+-}+ F^a,\end{align}
where $E^a=E^{a1}+E^{a2}$, $B^a=B^{a1}+iB^{a2}$, the solution to \eqref{eq:light-cone coordinate BPS 1}, \eqref{eq:light-cone coordinate BPS 2}, is \eqref{eq:ET solution 1}, \eqref{eq:ET solution 2}, \eqref{eq:ET solution 3}, with
\begin{align}a=b=\frac{1}{\sqrt{2}},\ \ \ \ \theta=-\kappa=-\frac{\pi}{4}. \end{align}
The 't Hooft Polyakov solution does not satisfy the equation \eqref{eq:light-cone coordinate BPS 4}, but the equations of motion change once the fields are assumed static, and this changes the saturation equations to \eqref{eq:light-cone coordinate BPS 1}, \eqref{eq:light-cone coordinate BPS 2}, \eqref{eq:light-cone coordinate BPS 3}.

The equations \eqref{eq: BPS component 1}, \eqref{eq: BPS component 2} have been chosen to be in light-cone gauge, and the 't Hooft Polyakov solutions do not satisfy them, because the 't Hooft Polyakov solutions are foliated in equal-time.  To find the light-cone gauge solutions the static 't Hooft monopole can be boosted to the spatial light-cone coordinates $x,\bar{x},x^-$ that are foliated in $x^+$, or the light-front.  This is a singular boost, and the method for boosting the 't Hooft Polyakov monopole is unknown, because of the hyperbolic functions non-trivial limits.  A simpler way to find a solution foliated in light-cone is to use the boundary behavior for the vector field in the 't Hooft equations, or the Wu-Yang Monopole
\begin{align}&A^1_\mu= A^2_\mu= A^3_3= A^3_0=0,\\ \nonumber \\
&A^N_i=\frac{1}{\rho^2}(1-\cos(\theta))\epsilon_{ij}x_j,\\ \nonumber \\
&A^S_i=\frac{1}{\rho^2}(1+\cos(\theta))\epsilon_{ij}x_j,\end{align}
where $\rho$ is the transverse radius $\rho^2=x_1^2+x_2^2$. Boosting the Wu-Yang monopole along the third axis and taking the limit of the rapidity to $\frac{\pi}{4}$ yields the ultra relativistic monopole \cite{24} \footnote{Equation \eqref{eq:relativistic monopole} is related to (4), (5) in \cite{24} by the redefinition of the step function found in appendix \ref{sec:a}.}
\begin{equation}A_i=\epsilon^{ij}\frac{x^j}{\rho^2}\theta(x^-).\label{eq:relativistic monopole}\end{equation}
The scalars are invariant under the boost;  thus, the scalar solution must be derived from the ultra-relativistic vector field.  If  $\Phi^a$ has one component 

\begin{equation}\Phi^1=0,\Phi^2=0,\Phi^3=\Phi,\end{equation}
this reduces the equations \eqref{eq: BPS component 1}, \eqref{eq: BPS component 2} to
\begin{align}&\partial^+A=\frac{1}{4}\partial \Phi,\label{eq:light-cone BPS 1} \\ \nonumber \\
&\partial\bar{A}=-\frac{1}{8}\partial^+\Phi,\label{eq:light-cone BPS 2}\end{align}
where $A=A_1+iA_2$.  Since bare distribution functions without integrals are not well defined, the solutions are integrated over a distance $L$;  therefore, the solutions to \eqref{eq:light-cone BPS 1}, \eqref{eq:light-cone BPS 2} are 
\begin{align}
&A=-i\frac{1}{4}\frac{x}{\rho^2}\int^{\frac{L}{2}}_{-\frac{L}{2}}dy^-\{\theta(y^--x^-)\},\label{eq:solution 1}\\ \nonumber \\
&\Phi=-i\ln(\rho)\int^{\frac{L}{2}}_{-\frac{L}{2}}dy^-\{\delta(y^--x^-)\}. \label{eq:solution 2}\end{align}
Substituting the previous solutions into the charge \eqref{eq:charge} yields zero 
\begin{equation}\int d^3x\{\partial_\nu F^{+\nu}\}=L\int dS\{Im(\frac{x}{\rho^2})\}=0,\end{equation}
where the surface is the transverse $S^1$.  This is a trait shared by all ultra-relativistic monopoles because of their pure gauge nature everywhere outside of the $x^-$ interval $(-\frac{L}{2},\frac{L}{2})$.

The Wu-Yang monopole singularities are not present in the 't Hooft Polyakov monopole in an equal-time theory.  It is unknown how to formulate a 't Hooft Polyakov equivalent solution in the light-cone theory, but hopefully it will solve some of the problems with the singularities at the origin in the solutions \eqref{eq:solution 1}, \eqref{eq:solution 2}, like the equal-time theory.  The solutions \eqref{eq:solution 1}, \eqref{eq:solution 2} are the only known solutions to the light-cone BPS theory. These solitons come from the static monopole boundary conditions that are boosted to the infinite momentum frame discussed for arbitrary gauge group in section \ref{sec:3.2}.  In the light-front frame these ultra-relativistic particles have mass equal to the electric charge that is zero for the solutions \eqref{eq:solution 1}, \eqref{eq:solution 2}.  

The analysis in this section so far has been based on solutions with the fermions set to zero.  The superfield formulation yields equations that generalize the BPS equations to include fermions.  These are the supersymmetric version of the BPS equations (SBPS)
\begin{align}&\epsilon^{mn}D^{ab}\frac{1}{\partial^+}\bar{\chi}^b_n=0, \\ \nonumber \\
&\partial^+A^a=\frac{1}{4}\epsilon^{mn}D^{ab}\bar{C}_{mn}^b+\frac{1}{\sqrt{2}}igf^{abc}\bar{\chi}^b_m\frac{1}{\partial^+}\bar{\chi}^c_n,\\ \nonumber \\
&\partial^+\bar{\chi}^{a}_m=i\frac{1}{3}(-\epsilon_{mnpq}\epsilon^{np}D^{ab}\chi^{bq}-gf^{abc}\epsilon^{np}(\bar{C}^{b}_{np}\bar{\chi}^c_m+\bar{C}^{b}_{mn}\bar{\chi}^c_p+\partial^+C^{cmn}\frac{1}{\partial^+}\bar{\chi}^c_p)),\\ \nonumber \\
&-\frac{1}{4}\epsilon_{mn}\partial^+C^{amn}=\nonumber\\ \nonumber \\
&=\frac{1}{\partial^+}(D^{ab}\partial^+\bar{A}^b)+\frac{1}{4}gf^{abc}(\frac{1}{\partial^+}(\bar{C}^b_{mn}\partial^+C^{c mn})+\frac{1}{\sqrt{2}}i(\frac{1}{\partial^+}\bar{\chi}^b_m\partial^+\chi^{cm}-3\bar{\chi}^b_m\chi^{cm})),\\ \nonumber \\
&\chi^{am}=\frac{1}{\sqrt{2}}gf^{abc}[\partial^{+}\bar{A}^b\frac{1}{\partial^+}\bar{\chi}^c_n-\frac{1}{\partial^+}\chi^{bp}\bar{C}^c_{np}].
\end{align}
These generalized BPS equations describe a non-linear field theory of particles with mass equal to the electric charge which is foliated in the light-cone frame, and are the supersymmetric generalization of the ultra-relativistic solution \eqref{eq:solution 1}, \eqref{eq:solution 2}.  In superspace the solutions to the previous equations make up a supersoliton $\tilde{\phi}^a$ that satisfies \eqref{eq:Superfield BPS}. In this article, there will not be an attempt to find the solutions to the SBPS equations.  These equations are presented to note that the bosonic theory is a subset of the arbitrary theory given by the previous equations, because equations \eqref{eq: BPS component 1}, \eqref{eq: BPS component 2} can be obtained by setting the fermions to zero in the SBPS equations.

\section{Conclusion}

\ \ \ \ In this article the BPS theory for the $\mathcal{N}=4$ light-cone superspace was derived, and a superfield equivalent of the Bogomol'nyi equations is the main result.  The difference between the light-cone superspace BPS theory and the normal BPS calculations is the light-cone includes fermions, and is the first superspace formulation of the BPS bound.  The simplicity of the light-cone chiral superfield with the inside-out constrant allows for the field theories bound to be formulated with the same expression as the algebra, the only difference being the algebraic operators are switched with infinitesimal transformations.  Since, the quadratic form is the fundamental feature that defines BPS theories in light-cone superspace it would seem that all other superspace theories should have a quadratic form;  although, it is unknown how to formulate the quadratic form in non-maximal supersymmetric theories or equal-time superspace.

These light-cone BPS theories are interpreted as the static equal-time monopole boosted to the ultra-relativistic solution;  although, it would be interesting to find solutions that have non-zero mass and fermions.  Furthermore, it would be interesting to construct an equal-time generalization of the BPS equations that includes fermions.   These formulas should match the conclusions of this article, and could be constructed from the SBPS equations presented.  The most interesting use of this article would be to construct a supermanifold moduli space for the superspace BPS equation presented.  This space would be the supersymmetric generalization of the hyperkahler Atiyah and Hitchin manifold.

The calculations presented can be extended to many different theories.  The recently formulated BLG theory in light-cone superspace is not a good candidate, since it has no known central charges.  A simple reduction of the $\mathcal{N}=4$ superfield to its $\mathcal{N}=2$ version would also be an interesting calculation.  It is unknown whether theories without maximal supersymmetry like $\mathcal{N}=2$ have a quadratic form.  It would be surprising if it did not have this feature, since the theory should have a BPS formulation similar to the $\mathcal{N}=4$.  Lastly, maximal supergravity theories superspace BPS formalism can be found using the quadratic form, although the usefulness of these equations is unknown at the current time.

\section{Acknowledgements}

\ \ \ \ I would like to acknowledge my advisor Pierre Ramond for all his valuable discussions.  Charles Sommerfield told me that the solutions could be constructed out of a Dyon boosted to the infinite momentum frame before any calculations had started.  Dmitry Belyaev read over the manuscript and gave me alot of advice on how to convey the dimensional reduction techniques. Sung-Soo Kim read over the manuscript, and has always offered an in-depth analysis of my project.  This research is partially supported by the Department of Energy Grant No. DE-F60297ER41C29.
\appendix
\section{Conventions}
\label{sec:a}
\ \ \ \ In four dimensions denoted by index $\mu=0,1,2,3$ the longitudinal and time coordinates can be arranged into the two light-cone lines

\begin{equation}x^\pm=\frac{1}{\sqrt{2}}(x_0\pm x_{3}),\ \ \ \ \partial^\pm=\frac{1}{\sqrt{2}}(\partial_0\pm\partial_{3}).\end{equation}
The light-cone metric is off diagonal in the plus minus coordinates $g^{+-}=g^{-+}=-1$.  The rest of the coordinates are Euclidian and are denoted by $i=1,2$, $g^{ij}=\delta^{ij}$.  

The light-cone gauge condition makes the temporal component of the vector field zero: $A^+=0$.  The minus coordinate $A^-$ is eliminated using the equations of motion.  To solve for the minus component of the vector field we must integrate over the derivative $\partial^+$, and introduce non-localities.  The greens function for the minus coordinate $x^-$ can be used to solve for this dependent field

\begin{equation}\partial^+G(x^-,y^-)=\delta(x^--y^-).\end{equation}
The previous equation is just the equation for a step function: $G(x^-,y^-)=\frac{1}{2}\theta(x^--y^-)$, where

\begin{align}&\theta(x^--y^-)=1\ for\ x^->y^-,\nonumber\\ \nonumber \\
&\theta(x^--y^-)=0\ for\ x^-=y^-,\nonumber\\ \nonumber \\
&\theta(x^--y^-)=-1\ for\ x^-<y^-;\end{align}
thus, when integrating over $\partial^+$ it is denoted by $\frac{1}{\partial^+}$, and has the meaning

\begin{equation}\frac{1}{\partial^+}f(x^-)=\frac{1}{2}\int dy^-\{\theta(x^--y^-)f(y^-)\}.\end{equation}
There is a more sophisticated prescription found in \cite{25}, but for the purposes of this article this will suffice.

In this article four, six, and ten dimensions will be used.  Starting with the Pauli matrices

\begin{equation}\sigma^1=\left(\begin{array}{cc}0 & 1 \\1 & 0\end{array}\right),\ \ \ \sigma^2=\left(\begin{array}{cc}0 & -i \\i & 0\end{array}\right),\ \ \ \sigma^3=\left(\begin{array}{cc}1 & 0 \\0 & -1\end{array}\right),\end{equation}
the light-cone matrices are
\begin{align}&\sigma^+=\frac{1}{\sqrt{2}}(I_2+\sigma_3)=\sqrt{2}\left(\begin{array}{cc}1 & 0 \\0 & 0\end{array}\right),\\ \nonumber \\
&\sigma^-=\frac{1}{\sqrt{2}}(I_2-\sigma_3)=\sqrt{2}\left(\begin{array}{cc}0 &0  \\0 & 1\end{array}\right),\\ \nonumber \\
&\sigma=\frac{1}{\sqrt{2}}(\sigma^1+i\sigma^2)=\sqrt{2}\left(\begin{array}{cc}0 & 1\\ 0& 0\end{array}\right).\end{align}
These matrices have the properties: $(\sigma^\pm)^2=\sqrt{2}\sigma^\pm$, $\sigma^\pm\sigma^\mp=0$, $\sigma\bar{\sigma}=\sqrt{2}\sigma^+$, $\bar{\sigma}\sigma=\sqrt{2}\sigma^-$.  

The four dimensional Clifford algebra satisfies $\{\gamma^\mu,\gamma^\nu\}=-2g^{\mu\nu}$, and has the equal-time representation

\begin{align}&\gamma^0=\left(\begin{array}{cc}0 & I_2 \\ I_2 & 0\end{array}\right),\ \ \ \gamma^x=\left(\begin{array}{cc}0 & \sigma^x \\ -\sigma^x & 0\end{array}\right),\label{eq:equal-time}\\ \nonumber \\
&\gamma_5=i\gamma_0\gamma_1\gamma_2\gamma_3=\left(\begin{array}{cc}-I_2 & 0 \\0 & I_2\end{array}\right),\ \ \ \ C_4=\left(\begin{array}{cccc}0 & 1 & 0 & 0 \\-1 & 0 & 0 & 0 \\0 & 0 & 0 & -1 \\0 & 0 & 1 & 0\end{array}\right),\end{align}
where $x=1,2,3$.  The matrices $\gamma_5$ are used for the four dimensional Weyl constraint $\psi=\pm\gamma_5\psi$, and $C_4$ is used for the Majorana constraint $\bar{\psi}=\psi^TC_4$, for any four dimensional spinor $\psi$, where $\bar{\psi}=\psi^T\gamma^0$.  The minus sign in the Weyl condition is the anti-Weyl spinor constraint.  The matrices \eqref{eq:equal-time} can be used to construct a Clifford algebra for the light-cone metric

\begin{align}&\gamma^\pm=\left(\begin{array}{cc}0 & \sigma^\pm \\ \sigma^\mp & 0\end{array}\right),\ \ \ \gamma^i=\left(\begin{array}{cc}0 & \sigma^i \\ -\sigma^i & 0\end{array}\right),\\ \nonumber \\
&\gamma=\frac{1}{\sqrt{2}}(\gamma^1+i\gamma^2)=\left(\begin{array}{cc}0 & \sigma \\ -\sigma & 0\end{array}\right).\end{align}

From four dimensions the higher dimensional gamma matrices can be defined.  One can construct the ten dimensional matrices denoted by $\Gamma^M$, where $M=0,1,2,...,9$, by starting with the first four dimensions

\begin{equation}\Gamma^\pm=i\gamma^\pm\otimes I_8,\ \ \ \Gamma=i\gamma\otimes I_8.\label{eq:matrices}\end{equation}
The other six dimensions are denoted by $I=4,5,6,7,8,9$.  These six extra ten dimensional gamma matrices can be found from a six dimensional Clifford algebra constructed with the 't Hooft symbols

\begin{equation}\eta_{ymn}=\epsilon_{ymn4}+\delta_{ym}\delta_{n4}-\delta_{yn}\delta_{m4},\end{equation}

\begin{equation}\tilde{\eta}_{ymn}=\epsilon_{ymn4}-\delta_{ym}\delta_{n4}+\delta_{yn}\delta_{m4},\end{equation}
where $m,n=1,2,3,4$ and the $\epsilon_{mnpq}$ symbol is a Levi-Civita tenser defined by the permutation group of four objects;  furthermore, the objects

\begin{align}&\Sigma_{Imn}=\eta_{ymn}\delta^I_y+i\tilde{\eta}_{ymn}\delta^I_{y+3},\\ \nonumber \\
&\Sigma^{Imn}=\eta_{ymn}\delta^I_y-i\tilde{\eta}_{ymn}\delta^I_{y+3},\end{align}
yield the matrices

\begin{equation}\tilde{\Gamma}^I=\left(\begin{array}{cc}0 & \Sigma^{Imn} \\ \Sigma_{Imn} & 0\end{array}\right),\end{equation}
that satisfy the Clifford algebra $\{\Gamma^I,\Gamma^J\}=-2g^{IJ}$, where $g^{IJ}$ is the six dimensional metric.  A complete list of the properties of the t'Hooft matrices can be found in \cite{6}.  Now the final ten dimensional gamma matrices are

\begin{equation}\Gamma^I=i\gamma_5\otimes\tilde{\Gamma}^I.\end{equation}
Finally, the ten dimensional matrices satisfy the Clifford algebra

\begin{equation}\{\Gamma^M,\Gamma^N\}=2g^{MN},\end{equation}
where $g^{MN}$ is the ten dimensional metric, and the Weyl, Majorana constraints are defined with the matrices

\begin{align}&\Gamma_{11}=i\Gamma_0...\Gamma_9=\gamma_5\times \left(\begin{array}{cc} I_4 & 0 \\0 & -I_4\end{array}\right),\\ \nonumber \\
&C=iC_4 \otimes \left(\begin{array}{cc}0 & I_4 \\I_4 & 0\end{array}\right),\end{align}
respectively.


\begin{thebibliography}{99} 
\bibitem{1}L. Brink, O. Lindgren and B. E. W. Nilsson, ÒThe Ultraviolet Finiteness Of The N=4 
Yang-Mills Theory,Ó Phys. Lett. B123, 323 (1983). 
\bibitem{2}S. Mandelstam, ÒLight Cone Superspace And The Ultraviolet Finiteness Of The N=4 
Model,Ó Nucl. Phys. B213, 149 (1983). 
\bibitem{3}Lars Brink, John H. Schwarz and J. Scherk, Supersymmetric Yang-Mills Theory, December 1976 
\bibitem{4} H. Osborn, Phys. Lett. 83B (1979) 321.
\bibitem{5} Lars Brink, Olof Lindgren, Bengt E.W. Nilsson, Jun 1982. 23pp. Nucl.Phys.B212:401,1983.
\bibitem{6} Dmitry V. Belyaev, Dynamical Supersymmetry in Maximally Supersymmetric Gauge Theories, October 2009,  \href{http://arxiv.org/abs/0910.5471}{hep-th/0910.5471}.
\bibitem{7} M. K. Prasad and C. H. Sommerfield, Phys. Rev. Lett. 35 (1975) 760.
\bibitem{8}  E. B. BogomolÕnyi, Sov. J. Nucl. Phys. 24 (1976) 449.
\bibitem{9} E. Witten and D. Olive, Phys. Lett. 78B (1978) 97.
\bibitem{10} Jeffrey A. Harvey, Magnetic Monopoles, Duality, and Supersymmetry,  March 1996.  \href{http://arxiv.org/abs/hep-th/9603086}{hep-th/9603086}
\bibitem{11} Paolo Di Vecchia, Duality in supersymmetric N=2,4 gauge theories, Mach 1998, \href{http://arxiv.org/abs/hep-th/9803026}{arXiv: hep-th/9803026v2}
 
\bibitem{12}P. A. M. Dirac, ÒForms Of Relativistic Dynamics,Ó Rev. Mod. Phys. 21, 392 (1949). 
\bibitem{13} Sudarshan Ananth, Lars Brink, Sung-Soo Kim, Pierre Ramond,Non-linear Realization of 
$PSU(2,2 |4)$ on the Light-Cone, May 2005. 29pp, \href{http://arxiv.org/abs/hep-th/0505234}{hep-th/0505234}
\bibitem{14} Pierre Ramond, Still in Light-Cone Superspace,  October 2009, \href{http://arxiv.org/abs/0910.1993}{arXiv: 0910.1993.}
\bibitem{15} Dmitry Belyaev, Lars Brink, Sung-Soo Kim, Pierre Ramond, The BLG Theory in Light-Cone Super-Space, January 2010, \href{http://arxiv.org/abs/1001.2001}{arXiv:1001.2001.}\\\\
\bibitem{16} Dmitry Belyaev, Mass-Deformed BLG Theory in Light-Cone Superspace, June 2010, \href{http://arxiv.org/abs/1006.1646}{arXiv:1006.1646v1}
\bibitem{17} Sudarshan Ananth, Lars Brink, Rainer Heise and Harald G. Svendsen, The N=8 Supergravity Hamiltonian as a Quadratic Form, \href{http://arxiv.org/abs/hep-th/0607019}{arXiv 06070192}\\\\
\bibitem{18}M. F. Atiyah and N. J. Hitchin, Phys. Lett. 107A (1985) 21. 
\bibitem{19}Lars Brink, Anna Tollsten, N=4 Yang-Mills Theory In Terms Of N=3 And N=2 Light Cone Superfields, May 1984, Nucl.Phys.B249:244,1985.
\bibitem{20}Nathan Seiberg, Notes on Theories with 16 Supercharges, May 1997, \href{http://arxiv4.library.cornell.edu/abs/hep-th/9705117v1}{arXiv:hep-th/9705117v2}
\bibitem{21}Nicholas Dorey, Christophe Fraser, Timothy J. Hollowood, Marco A. C. Kneipp,  S-duality in N=4 supersymmetric gauge theories with arbitrary gauge group, May 1996,  \href{http://arxiv.org/abs/hep-th/9605069}{arXiv:hep-th/9605069v1}
\bibitem{22}G. 't Hooft, Nucl. Phys. B79 (1974) 276. 
\bibitem{23}A.M. Polyakov, JETP Lett. 20 (1974) 194. 
\bibitem{24}Saurya Das, Parthasarathi Majumdar, Charge-monopole versus Gravitational Scattering at Planckian Energies, July 1993 , \href{http://arxiv.org/abs/hep-th/9307182v1}{arXiv:hep-th/9307182v1}
\bibitem{25}S. Mandelstam, Nucl. Phys. B 213 (1983) 149.
\end{thebibliography}
\end{document}